\documentclass[twocolumn]{aastex62}
\usepackage{float,graphicx,amsmath,multirow}
\usepackage{color}
\usepackage[version=4]{mhchem}
\usepackage{booktabs}
\usepackage{gensymb}

\begin{document}

\title{Early Science from GOTHAM: Project Overview, Methods, and the Detection of Interstellar Propargyl Cyanide (\ce{HCCCH2CN}) in TMC-1}
\author{Brett A. McGuire}
\affiliation{Department of Chemistry, Massachusetts Institute of Technology, Cambridge, MA 02139, USA}
\affiliation{National Radio Astronomy Observatory, Charlottesville, VA 22903, USA}
\affiliation{Center for Astrophysics $\mid$ Harvard~\&~Smithsonian, Cambridge, MA 02138, USA}
\author{Andrew M. Burkhardt}
\affiliation{Center for Astrophysics $\mid$ Harvard~\&~Smithsonian, Cambridge, MA 02138, USA}
\author{Ryan A. Loomis}
\affiliation{National Radio Astronomy Observatory, Charlottesville, VA 22903, USA}
\author{Christopher N. Shingledecker}
\affiliation{Center for Astrochemical Studies, Max Planck Intitute for Extraterrestrial Physics, Garching, Germany}
\affiliation{Institute for Theoretical Chemistry, University of Stuttgart, Stuttgart, Germany}
\author{Kin Long Kelvin Lee}
\affiliation{Center for Astrophysics $\mid$ Harvard~\&~Smithsonian, Cambridge, MA 02138, USA}
\author{Steven B. Charnley}
\affiliation{Astrochemistry Laboratory and the Goddard Center for Astrobiology, NASA Goddard Space Flight Center, Greenbelt, MD 20771, USA}
\author{Martin A. Cordiner}
\affiliation{Astrochemistry Laboratory and the Goddard Center for Astrobiology, NASA Goddard Space Flight Center, Greenbelt, MD 20771, USA}
\affiliation{Institute for Astrophysics and Computational Sciences, The Catholic University of America, Washington, DC 20064, USA}
\author{Eric Herbst}
\affiliation{Department of Chemistry, University of Virginia, Charlottesville, VA 22904, USA}
\affiliation{Department of Astronomy, University of Virginia, Charlottesville, VA 22904, USA}
\author{Sergei Kalenskii}
\affiliation{Astro Space Center, Lebedev Physical Institute, Russian Academy of Sciences, Moscow, Russia}
\author{Emmanuel Momjian}
\affiliation{National Radio Astronomy Observatory, Socorro, NM 87801, USA}
\author{Eric R. Willis}
\affiliation{Department of Chemistry, University of Virginia, Charlottesville, VA 22904, USA}
\author{Ci Xue}
\affiliation{Department of Chemistry, University of Virginia, Charlottesville, VA 22904, USA}
\author{Anthony J. Remijan}
\affiliation{National Radio Astronomy Observatory, Charlottesville, VA 22903, USA}
\author{Michael C. McCarthy}
\affiliation{Center for Astrophysics $\mid$ Harvard~\&~Smithsonian, Cambridge, MA 02138, USA}

\correspondingauthor{Brett A. McGuire}
\email{brettmc@mit.edu}

\begin{abstract}

We present an overview of the GOTHAM (GBT Observations of TMC-1: Hunting Aromatic Molecules) Large Program on the Green Bank Telescope.  This and a related program were launched to explore the depth and breadth of aromatic chemistry in the interstellar medium at the earliest stages of star formation, following our earlier detection of benzonitrile ($c$-\ce{C6H5CN}) in TMC-1.  In this work, details of the observations, use of archival data, and data reduction strategies are provided.  Using  these observations, the interstellar detection of propargyl cyanide (\ce{HCCCH2CN}) is described, as well as the accompanying laboratory spectroscopy.  We discuss these results, and the survey project as a whole, in the context of investigating a previously unexplored reservoir of complex, gas-phase molecules in pre-stellar sources.  A series of companion papers describe other new astronomical detections and analyses.

\end{abstract}

\section{Introduction, Overview, and Motivation}
\label{intro}

More than 204 individual molecular species have been detected to date in the interstellar medium (ISM; \citealt{McGuire:2018mc}). Of these, only 10 ($<$5\%) are fully saturated - meaning they have only single bonds.  It is perhaps then surprising that few aromatic molecules, which are greatly stabilized by the presence of delocalized electrons shared across their bonds, have been seen in the ISM.  Indeed, despite their dominant place in terrestrial organic chemistry \citep{Lipkus:2008dd,Ruddigkeit:2012mh}, only two (non-fullerene) benzene-containing molecules have been detected: benzene itself ($c$-\ce{C6H6}; \citealt{Cernicharo:2001mw}) and, recently, benzonitrile (cyanobenzene; $c$-\ce{C6H5CN}; \citealt{McGuire:2018it}).  

The few detections of interstellar benzene have all been toward post-AGB/pre-planetary nebula sources such as CRL 618 and SMP-LMC 11 \citep{Cernicharo:2001mw,Malek:2012kj,Kraemer:2006va,GarciaHernandez:2016tn}.  This finding is consistent with the broadly discussed theory of the formation of large aromatic molecules, such as the polycyclic aromatic hydrocarbons (PAHs), in the circumstellar envelopes of soot-producing, post-AGB stars: ``top-down chemistry."  PAHs can then be broken down in the harsh environment until they eventually are ejected from the region with some observed distribution that includes benzene and other small benzene-containing species (see \citealt{Tielens:2008fx} and references therein for an extensive discussion).  

Benzonitrile, however, was detected in the pre-stellar source TMC-1.  The presence of this benzene-ring molecule in a pre-stellar source, nearly as far separated from the post-AGB phase as is possible, is therefore surprising.  If the predominant source of benzene is indeed from post-AGB molecular synthesis, this would likely imply that at least a portion of the chemical inventory of this dark cloud was inherited from a previous generation of stars. Yet, it is also possible that a substantial population of benzene can be formed via a ``bottom-up chemistry" from smaller organic precursors present in the cloud.  Indeed, recent laboratory work has shown that plausible formation pathways to benzene and benzonitrile may exist in sources such as TMC-1 \citep{Jones:2011yc,Balucani:1999it,Lee:2019dh,Cooke:2020we}.  Regardless of the relative dominance of these formation scenarios -- and it is likely some combination of the two -- the unexpected presence of large aromatic molecules in pre-stellar sources raises a number of questions.  
\begin{enumerate}
    \item Are there other aromatic species beyond benzonitrile (and by proxy, benzene) in TMC-1?  
    \item What other precursors are available in TMC-1 to build these aromatic molecules in a bottom-up scenario?
    \item Is this chemistry unique to TMC-1, or is it widespread throughout the early star-formation process?
    \item Do these species survive (and thus impact) the collapse to, and formation of, a protostar and the simultaneous chemical evolution?
\end{enumerate}
To begin to address these questions, two observing programs have been undertaken with the Robert C. Byrd 100\,m Green Bank Telescope (GBT) to explore more fully both the aromatic chemistry in TMC-1 specifically and at the earliest stages of the star and planet formation processes more generally. 

The first program, GBT Observations of TMC-1: Hunting for Aromatic Molecules (GOTHAM), is a large-scale high-spectral resolution, high-sensitivity, large-bandwidth spectral line survey of TMC-1.  The primary goals of GOTHAM are to address the first two questions by establishing the chemical inventory in this source by performing a very high sensitivity wide-band spectral line survey,  and then use this information to explore, through laboratory and modeling work, bottom-up chemistry in TMC-1. The second project, A Rigorous K-band Hunt for Aromatic Molecules (ARKHAM), is a high-sensitivity search for benzonitrile in pre-stellar and proto-stellar sources outside of TMC-1, addressing the second two questions.

A series of six papers provides early science results from the ongoing GOTHAM and ARKHAM programs. This paper describes the motivation behind GOTHAM, the observing strategy, calibration, and reduction of the survey data, along with the interstellar detection of propargyl cyanide (\ce{HCCCH2CN}).  \citet{Burkhardt:2020aa} presents the first results of ARKHAM: detection of benzonitrile in four additional sources, a finding which demonstrates the widespread existence of aromatic chemistry throughout the earliest stages of star formation. \citet{Loomis:2020aa} provides a detailed description of the fitting and velocity stacking techniques used to analyze the data, and applies these techniques to the detection of \ce{HC11N} in TMC-1, a long chain polyyne the detectability of which has been the subject of recent debate in the literature \citep{Bell:1997vj,Loomis:2016js,Cordiner:2017dq}.   \citet{McGuire:2020aa} presents the detections of both 1-~and~2-cyanonaphthalene (\ce{C10H7CN}), the first individual PAHs detected in the ISM, in TMC-1. \citet{McCarthy:2020aa} describes  detection of 1-cyano-cyclopentadiene ($c$-\ce{C5H5CN}), a highly polar five-membered ring, in the same source.  Finally, \citet{Xue:2020aa} describes the astronomical discovery of \ce{HC4NC}, the isocyanide isomer of the commonly observed cyanopolyyne \ce{HC5N}, and explores the implications for the formation pathways of these widespread molecules in the ISM.

\section{TMC-1 Properties}

The Taurus Molecular Cloud is a nearby (140 pc; \citealt{Onishi:2002vr}), well-studied molecular cloud complex in which 34 molecules have been detected for the first time in the ISM \citep{McGuire:2018mc}.\footnote{This series of papers brings that total to 40, or almost 20\% of all known interstellar species.}  The cyanopolyyne peak within this source has been shown to be particularly  rich in molecular species, and is one of the reasons TMC-1 is often used as the prototypical molecular dark cloud, especially for benchmarking astrochemical models of the early evolution of chemistry before cloud collapse \citep{Agundez:2013ga}. {The chemistry of the cyanopolyyne peak in particular has been of substantial interest in recent years \citep{Fuente:2019cw,Gratier:2016fj}.  As well, the Taurus Molecular cloud itself has seen sustained interest, especially in the areas of chemical evolution \citep{Scibelli:2020jr}.}  Observations at a variety of wavelengths and multiple tracers suggest the effective size of the molecular emitting region is $\sim$20--40$^{\prime\prime}$ (see, e.g., \citealt{Feher:2016gf} and references therein).

Because it is so quiescent, TMC-1 is characterized by extremely low rotational excitation temperatures ($T_{ex}$~=~5--10~K) and very narrow linewidths.  \citet{Kaifu:2004tk} reported widths of $\sim$0.4~km~s$^{-1}$ FWHM at a resolution that ranged from  0.22 to 1.26\,km s$^{-1}$.  Observations at higher spectral resolution ($\leq$0.05\,km~s$^{-1}$)  both in the GOTHAM project \citep{Loomis:2020aa} {and in other recent work (e.g., \citealt{Dobashi:2018kd,Dobashi:2019ev})} found that lines of most species consist of multiple velocity components with widths of $<$0.2~km~s$^{-1}$.  For this work, and those presented in the other GOTHAM papers, we adopt a uniform molecular hydrogen column density of $N_T(\ce{H2})$=~=~$10^{22}$~cm$^{-2}$ and a molecular hydrogen density of $n(\ce{H2})$~=~$2\times10^4$~cm$^{-3}$ toward the TMC-1 cyanopolyyne peak from the work of \citet{Cernicharo:2018bv}.

\section{Observations}
\label{observations}

This paper describes the GOTHAM Large Project.  Details of the  ARKHAM project are provided in \citet{Burkhardt:2020aa}.  The GOTHAM observations presented here were carried out between February 2018 and May 2019 on the Robert C. Byrd 100-m Green Bank Telescope in Green Bank, West Virginia under project codes GBT18A\_333 and GBT18B\_007. The target was TMC-1 at (J2000) $\alpha$~=~04$^h$41$^m$42.50$^s$ $\delta$~=~+25$^{\circ}$41$^{\prime}$26.8$^{\prime\prime}$.  Pointing and focus observations were performed using J0530+1331 as the calibrator source at the beginning of each observing session, and every subsequent 1--2~hours, depending on the weather.  Typical pointing solutions converged to $\lesssim$5$^{\prime\prime}$.  Observations were performed in ON-OFF position-switched mode with 2 minutes on target and 2 minutes off target, and an off position throw of 1$^{\circ}$.  In addition to these new observations, we have also used the data acquired for the original benzonitrile detection under project codes GBT17A-164 and GBT17A-434.  The detailed observing strategy for these archival data is outlined in \citet{McGuire:2018it}, but is largely identical to that used for the GOTHAM data.  The archival data were taken in their raw form from the archive and were re-calibrated and re-reduced uniformly with the new GOTHAM observations, to ensure consistency.

\subsection{Spectral Configuration}

\begin{figure*}[!ht!]
    \centering
    \includegraphics[width=\textwidth]{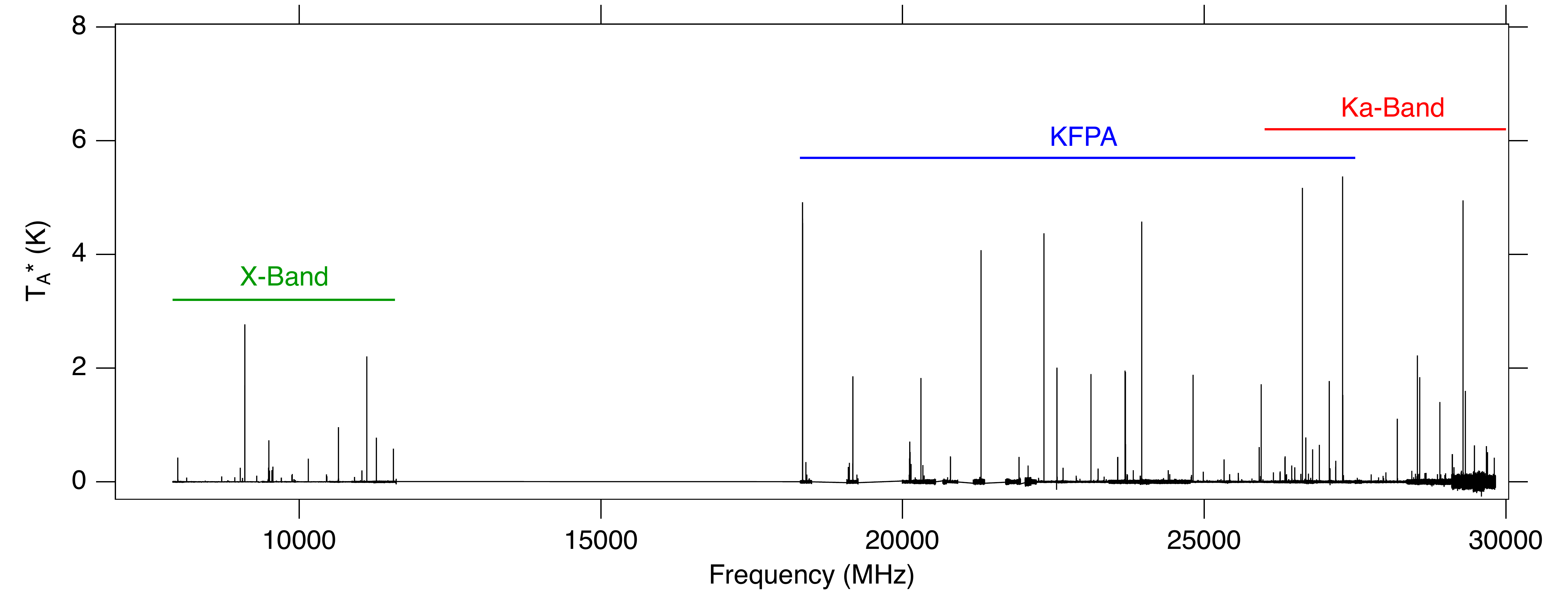}
    \caption{Current total coverage of the GOTHAM Large Project from the observations described here and archival observations described in \citet{McGuire:2018it}. The receivers used in each frequency range are labeled above the spectra.}
    \label{coverage}
\end{figure*}

The full spectra coverage of the survey is shown in Figure~\ref{coverage}.  Three receivers were used to cover the frequency range of the observations, the X-band (8.0--11.6\,GHz), K-band Focal Plane Array (KFPA; 18.0--27.5\,GHz), and Ka-band (26.0--39.5\,GHz) receivers.  The VEGAS spectrometer was the backend for all observations \citep{Roshi:2012he}.

\subsubsection{X-Band}

In X-band, eight VEGAS spectrometer banks were configured in their 187.5~MHz modes with 131072 channels corresponding to a resolution of 1.4~kHz (0.05~km~s$^{-1}$ at 9~GHz). All eight spectrometers were routed to the single beam of this dual-circular polarization receiver, and both polarizations were recorded.  The half-power beam width (HPBW) of the GBT is $\sim$80$^{\prime\prime}$ at 9~GHz.

\subsubsection{K-Band}

The KFPA is a seven-pixel focal plane array receiver operating from 18--27.5~GHz \citep{Morgan:2008kb}. For these observations, all eight spectrometers of VEGAS were routed to the central pixel/beam of the KFPA, and both polarizations were recorded.  The VEGAS spectrometer banks were configured in their 187.5~MHz modes with 131072 channels corresponding to a resolution of 1.4~kHz (0.02~km~s$^{-1}$ at 23~GHz). The data obtained for GOTHAM begin at 22.2~GHz; data below 22.2 GHz were obtained from the archive (GBT17A-164 and GBT17A-434).  The observing strategy for these archival data is described in detail in \citep{McGuire:2018it}, but they have been uniformly re-reduced as part of the larger dataset here.  The HPBW of the GBT at 23~GHz is $\sim$33$^{\prime\prime}$.

\subsubsection{Ka-band}

The Ka-band receiver is a dual-beam receiver operating from 26--39.5~GHz, with only a single linear polarization available per beam.  Due to limitations in the IF system, only four VEGAS spectrometers can be routed to a single beam, and only the single polarization for that beam is obtained, in this case, LL.  For these observations, the four VEGAS spectrometer banks were configured in their 187.5~MHz modes with 131072 channels corresponding to a resolution of 1.4~kHz (0.015~km~s$^{-1}$ at 28~GHz).  The HPBW of the GBT at 28~GHz is $\sim$27$^{\prime\prime}$.

\subsection{Calibration}

All three receivers are primarily calibrated by means of an internal noise diode, which we assume gives an absolute flux density calibration uncertainty of, at best, $\sim$30\%.  The noise diode in the X-band receiver was calibrated as recently as 2018 and referenced to Karl G. Jansky Very Large Array (VLA) flux density measurements (see \url{http://www.gb.nrao.edu/GBTCAL/}), and is therefore assumed to be better than 30\%.  We have taken several steps to improve both the absolute flux calibration of the KFPA and Ka-band measurements, and to ensure relative agreement between the two (and with X-band).  The pointing source, J0530+1331 was observed after every pointing and focus performed as part of the observations, using an identical spectral setup and position-switched cadence with VEGAS as would be used for the subsequent science target. Because J0530+1331 has shown long-term variability (as well as short-term variability of order $\sim$20\%; \citealt{Gorshkov:2016hy}), we obtained new VLA flux density measurements of the source.  

The VLA observations of J0530+1331 were carried out on 6 May 2019 at K-band (18--26.5\,GHz) for a total of 25\,min. The 8-bit samplers utilized in these observations delivered a total frequency coverage that spanned the range 24 to 26\,GHz. The absolute flux density scale was set by observing the calibrator J0521+1638 (3C138) and using the Perley-Butler\,2017 flux density scale standard \citep{Perley:2017gn}. The elevation of both J0530+1331 and 3C138 was near $65\degree$ during these observations. The calibrator source 3C138 is also known to exhibit variability, and through regular monitoring observations with the VLA between 2016 and 2019, it was found that the flux density of 3C138 has increased by 8\% compared to the values of the Perley-Butler\,2017 standard (R. Perley, private communication). Accounting for this variability in 3C138, we measure the flux density value of $1.19 \pm 0.03$\,Jy at 25\,GHz for J0530+1331 in May 2019.  This value was then used to calibrate the GBT flux scaling.

The KFPA noise diode internal calibration is more recent than that of the Ka-receiver, and so we first calibrated the KFPA measurements to flux densities measured with the VLA. Because there is spectral overlap between the KFPA and Ka-receiver, we then use spectral line observations of the $J=3-2$ transition of \ce{HCC^{13}CN} at 27181~MHz with both receivers to bring the Ka-receiver measurements into the same calibration as the KFPA. Because of the short-term variability in J0530+1331, and the differences between receivers, we still assume the calibration accuracy is only $\sim$20\%, and fold this uncertainty into all calculations. This is usually the dominant source of uncertainty in our measurements.  Nevertheless, we are confident that the relative calibration between receivers is much better than this number.

\section{Data Reduction}

Initial data processing and calibration was performed using \textsc{gbtidl}.  Each ON-OFF position switched scan pair was corrected for Doppler Tracking, calibrated to the internal noise diodes, and then placed on the atmosphere-corrected $T_A$* intensity scale \citep{Ulich:1976yt}.  When available (for X-band and the KFPA), both recorded polarizations were averaged together to improve the signal-to-noise ratio.  The spectra were then manually cleaned of RFI and artifacts. {Baselines were removed with a polynomial fit of order appropriate to the baseline ripple observed, and that typically ranged from 1--20.  In all cases, the continuum model was inspected to ensure that the line profiles, which are exceedingly narrow compared to the model, were not affected.}  Finally, a noise-weighted average was performed to arrive at the final spectrum.

When convolved to a uniform velocity resolution of 0.05~km~s$^{-1}$ across the spectrum (corresponding to the lowest-resolution data at X-band), the RMS noise level varies from $\sim$2--20~mK, dependent entirely upon the integration time achieved at that frequency as of 10 May 2019.  A uniform sensitivity of $\sim$2~mK (at 0.05~km~s$^{-1}$) is expected across the entire band at the completion of the project, although we make use of the full 1.4~kHz resolution when possible.  A fully reduced and calibrated dataset will be provided to the community upon completion of the entire survey. As there is no further data intended to be collected at X-band, that portion of the survey is considered complete and has been provided in its reduced form as Supplementary Information.

\section{Detection of Propargyl Cyanide}

The laboratory rotational spectrum of propargyl cyanide\footnote{Propargyl cyanide  has several alternate names in the chemical literature including 3-butynenitrile and 1-cyanoprop-2-yne.} was first reported by \citet{Jones:1982vg} up to 39\,GHz.  Later work by \citet{Demaison:1985td}, \citet{McNaughton:1988tk}, and \citet{Jager:1990ty} extended the measurements to 300\,GHz and included a determination of the \ce{^14N} nuclear hyperfine parameters.  The dipole moment components ($\mu_a = 2.87$\,D, $\mu_b = 2.19$\,D, $\mu = 3.61$\,D) were derived in \citet{McNaughton:1988tk}.  Based on these spectra, \citet{Lovas:2006ty} performed an astronomical search for propargyl cyanide in TMC-1 while investigating its structural isomers cyanoallene (\ce{CH2CCHCN}) and methylcyanoacetylene (\ce{CH3CCCN}).  Using the GBT, they set a 1$\sigma$, beam-averaged upper limit of $N_T$~=~$2.8\times10^{11}$~cm$^{-2}$ for the column density of propargyl cyanide  assuming $T_{ex}$~=~4\,K, a 5\,mK RMS noise level for the $4_{1,4}-3_{1,3}$ transition at 21249~MHz, and  a resolution of 6.1\,kHz, equivalent to 0.09~km~s$^{-1}$.  Because the present observations cover a substantially wider frequency range at much higher sensitivity and resolution, it was necessary to systematically re-measure  hyperfine-split transitions of propargyl cyanide in the laboratory up to 40\,GHz so as to better match the new, high-quality astronomical data.

\subsection{Laboratory Measurements}

To derive rest frequencies to the accuracy required for TMC-1, more than 110 hyperfine-split $a$- and $b$-type features of propargyl cyanide have been measured to 2\,kHz between 5 and 40\,GHz (Table~\ref{pcn:lines}).  These measurements were made with a Fourier transform microwave spectrometer in combination with a supersonic jet discharge source, the same technique recently used to extend the high-resolution rotational spectroscopy of benzonitrile at centimeter-wavelengths (\citealt{McGuire:2018it}).  Strong lines of propargyl cyanide were observed using a mixture of acetylene (0.4\%) and \ce{CH3CN} (0.1\%) heavily diluted in Ne, and applying a voltage of 1300\,V as the gas passes through the throat of the nozzle source. Relative to previous high-resolution work, lines originating from higher $J$,  and therefore higher frequency, have been observed in the present study.  By varying ten parameters in a standard A-reduced asymmetric top Hamiltonian \citep{Watson:1977vk} with hyperfine interactions -- the three rotational constants, the five quartic centrifugal distortion constants, and two tensor terms that describe the nitrogen hyperfine structure -- it was possible to achieve a fit rms (2.5\,kHz) that is comparable to the measurement uncertainty. From these best-fit constants (Table~\ref{pcn:constants}), the astronomically most intense hyperfine-split transitions of this species can now be calculated to 0.03~km~sec$^{-1}$ or better, equivalent to about 1/5 of the linewidth in TMC-1, or the width of a single channel in the astronomical survey.

\subsection{Observational Analysis}

Full details of the observational analysis method are provided in \citet{Loomis:2020aa}.  In short, we have first performed a Markov-Chain Monte Carlo fit to the strongly-detected \ce{HC9N} cyanopolyyne and to $c$-\ce{C6H5CN}.  We detect four distinct velocity components contributing to the overall signal for lines of these species, and derive velocities ($v_{lsr}$), column densities ($N_T$), and source sizes ($\theta_s$) simultaneously with an excitation temperature ($T_{ex}$) and linewidth ($\Delta V$) following the conventions of \citet{Turner:1991um} which include corrections for optical depth.  The values of $v_{lsr}$, $\theta_s$, $T_{ex}$, and $\Delta V$ are then used as priors for MCMC analyses of other species with fewer and/or weaker lines, with $T_{ex}$ and $\Delta V$ assumed to be the same for all velocity components.  \ce{HC9N} is used as a starting point for linear molecules while $c$-\ce{C6H5CN} is used for cyclic species. 

\begin{figure*}[!t]
    \centering
    \includegraphics[width=\textwidth]{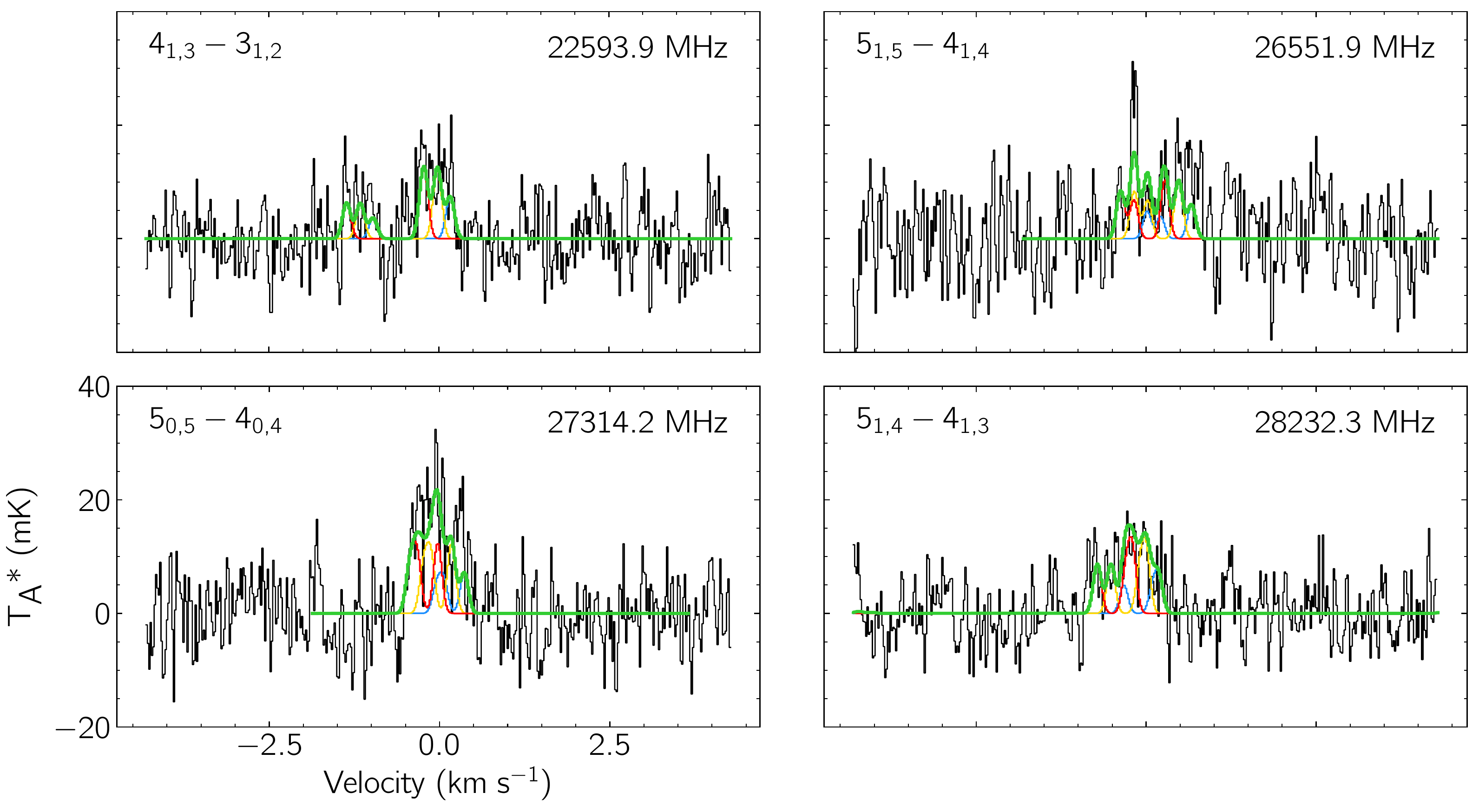}
    \caption{Individual line detections of propargyl cyanide in the GOTHAM data.  The spectra (black) are displayed in velocity space relative to 5.8\,km\,s$^{-1}$, and using the rest frequency given in the top right of each panel. Quantum numbers are given in the top left of each panel, neglecting hyperfine splitting. The best-fit model to the data, including all velocity components, is overlaid in green.  Simulated spectra of the individual velocity components are shown in: blue (5.615\,km\,s$^{-1}$), gold (5.804\,km\,s$^{-1}$), red (6.005\,km\,s$^{-1}$).  See Table~\ref{propargylcyanide_results}.}
    \label{spectra}
\end{figure*}

Using the priors from \ce{HC9N}, four lines belonging to propargyl cyanide were detected and fit above the present noise level of the observations: the $4_{1,3} - 3_{1,2}$, $5_{1,5} - 4_{1,4}$, $5_{0,5} - 4_{0,4}$, and $5_{1,4} - 4_{1,3}$ hyperfine-split transitions.  These are shown in Figure~\ref{spectra}.  More than 3700 transitions of propargyl cyanide fall within the range of GOTHAM's coverage, however, and contribute to the total flux seen for this molecule.  The MCMC analysis included all of these transitions (see Appendix \ref{app:lines}), and results in a significant detection of propargyl cyanide emission in three of the four velocity components in which \ce{HC9N} is found.  The resulting physical parameters, column densities, and excitation temperatures are given in Table~\ref{propargylcyanide_results}.

\begin{figure*}
    \centering
    \includegraphics[width=0.49\textwidth]{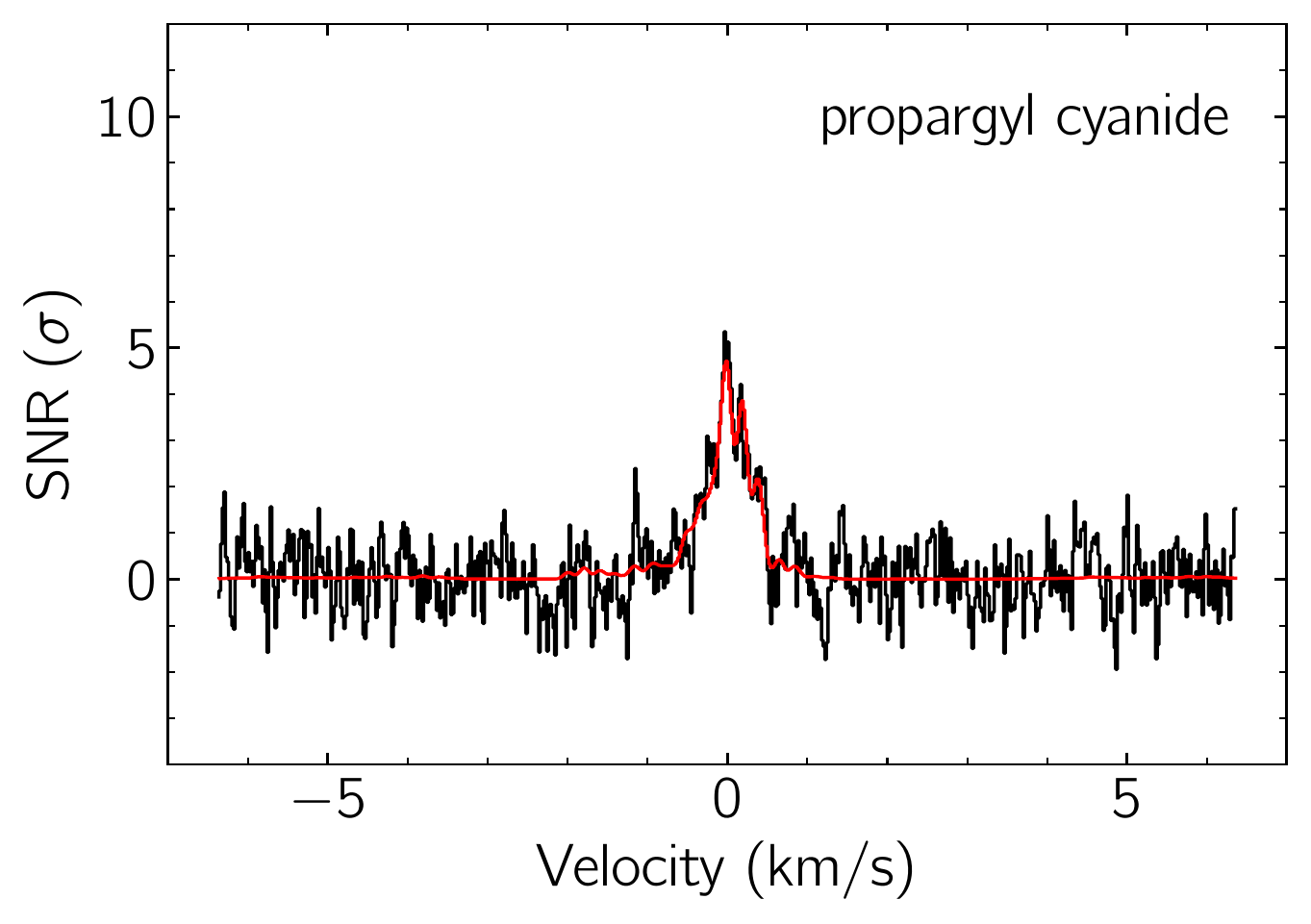}
    \includegraphics[width=0.49\textwidth]{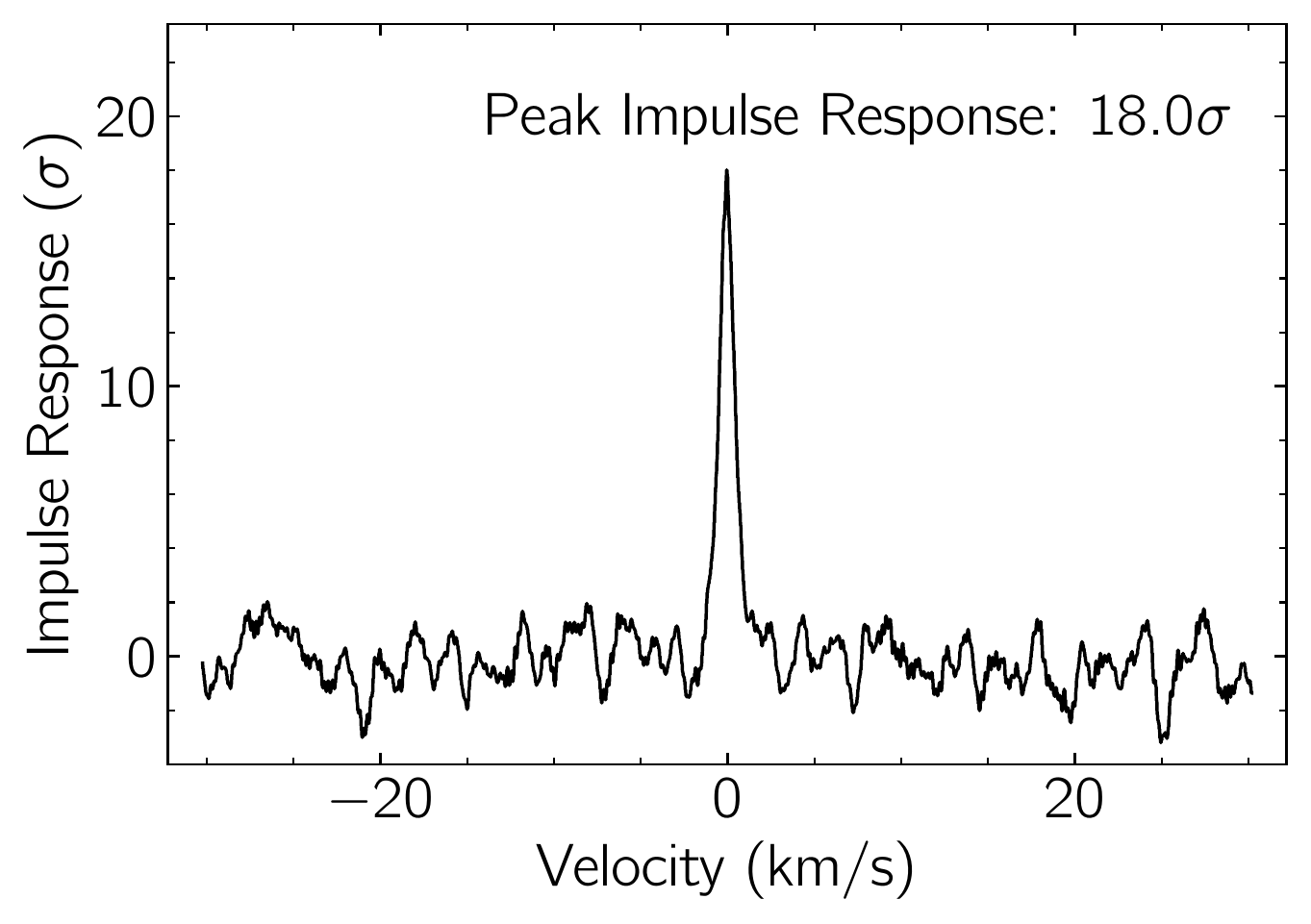}
    \caption{\emph{Left:} Velocity-stacked spectra of propargyl cyanide in black, with the corresponding stack of the simulation using the best-fit parameters to the individual lines in red.  The data have been uniformly sampled to a resolution of 0.02\,km\,s$^{-1}$.  The intensity scale is the signal-to-noise ratio of the spectrum at any given velocity. \emph{Right:} Impulse response function of the stacked spectrum using the simulated line profile as a matched filter.  The intensity scale is the signal-to-noise ratio of the response function when centered at a given velocity.  The peak of the impulse response function provides a minimum significance for the detection of 18.0$\sigma$.  See \citealt{Loomis:2020aa} for details.}
    \label{stack}
\end{figure*}

Using these parameters, we then perform an {intensity and noise-weighted} average (``stack") of both the observations and the best-fit model spectra in velocity-space, resulting in a substantial increase in signal-to-noise ratio (SNR) on a channel-by-channel basis.  This spectrum encapsulates the total flux of emission from propargyl cyanide contained within the bandwidth of GOTHAM, rather than only the flux coming from lines seen above the local noise level of the observations.  This is shown in Figure~\ref{stack} (left).   Finally, to determine the overall significance of the detection, we use the stacked model as a matched filter which is pushed through the stacked observations.  The resulting impulse response function represents the minimum statistical significance of the detection and is shown in Figure~\ref{stack} (right).

\begin{table*}
\centering
\caption{Propargyl cyanide best-fit parameters from MCMC analysis}
\begin{tabular}{c c c c c c}
\toprule
\multirow{2}{*}{Component}&	$v_{lsr}$					&	Size					&	\multicolumn{1}{c}{$N_T^\dagger$}					&	$T_{ex}$							&	$\Delta V$		\\
			&	(km s$^{-1}$)				&	($^{\prime\prime}$)		&	\multicolumn{1}{c}{(10$^{11}$ cm$^{-2}$)}		&	(K)								&	(km s$^{-1}$)	\\
\midrule
\hspace{0.1em}\vspace{-0.5em}\\
C1	&	$5.615^{+0.016}_{-0.016}$	&	$207^{+133}_{-122}$	&	$2.07^{+0.39}_{-0.31}$	&		&	\\
\hspace{0.1em}\vspace{-0.5em}\\
C2	&	$5.804^{+0.009}_{-0.010}$	&	$130^{+178}_{-84}$	&	$3.59^{+1.14}_{-0.44}$	&	$6.5^{+0.3}_{-0.3}$	&	$0.144^{+0.012}_{-0.010}$\\
\hspace{0.1em}\vspace{-0.5em}\\
C3	&	$6.005^{+0.008}_{-0.008}$	&	$192^{+131}_{-116}$	&	$3.56^{+0.53}_{-0.36}$	&		&	\\
\hspace{0.1em}\vspace{-0.5em}\\
\midrule
$N_T$ (Total)$^{\dagger\dagger}$	&	 \multicolumn{5}{c}{$9.21^{+1.31}_{-0.65}\times 10^{11}$~cm$^{-2}$}\\
\bottomrule
\end{tabular}

\begin{minipage}{0.75\textwidth}
	\footnotesize
	{Note} -- The quoted uncertainties represent the 16$^{th}$ and 84$^{th}$ percentile ($1\sigma$ for a Gaussian distribution) uncertainties.\\
	$^\dagger$Column density values are highly covariant with the derived source sizes.  The marginalized uncertainties on the column densities are therefore dominated by the largely unconstrained nature of the source sizes, and not by the signal-to-noise of the observations.  See Fig.~\ref{propargylcyanide_corner} for a covariance plot, and \citealt{Loomis:2020aa} for a detailed explanation of the methods used to constrain these quantities and derive the uncertainties.\\
	$^{\dagger\dagger}$Uncertainties derived by adding the uncertainties of the individual components in quadrature.
\end{minipage}

\label{propargylcyanide_results}

\end{table*}

\subsection{Astrochemical Modeling}

\begin{figure}
    \centering
    \includegraphics[width=0.49\textwidth]{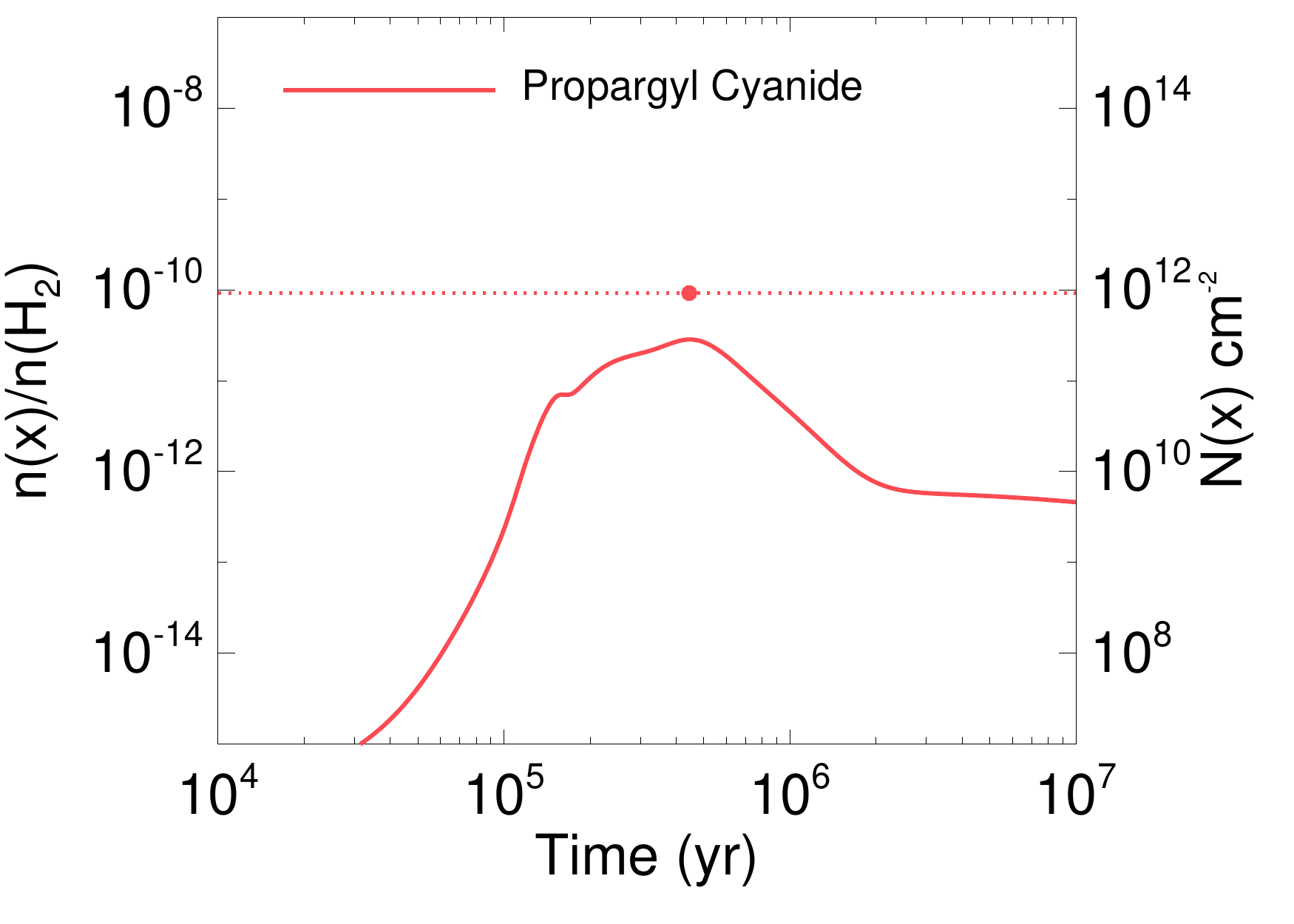}
    \caption{Calculated abundance of propargyl cyanide (solid line) in our TMC-1 simulation. Abundances from the MCMC analysis are represented by the dotted line and time of peak abundance by the filled circle. Note: observational errors given in Table 1 are not visible at the scale used.}
    \label{fig:modeling}
\end{figure}

In order to better understand the chemistry of propargyl cyanide in cold cores, we have simulated TMC-1 using astrochemical codes. Specifically, for this study we have used the \texttt{NAUTILUS}-v1.1 program \citep{ruaud_gas_2016} along with a modified version the KIDA 2014 network \citep{wakelam_2014_2015} also used in the analysis of the other species detected in the GOTHAM and ARKHAM surveys \citep{McGuire:2020aa,Loomis:2020aa,McCarthy:2020aa,Xue:2020aa,Burkhardt:2020aa}. Typical TMC-1 physical conditions were used, including $T_\mathrm{gas}=T_\mathrm{dust}=10$ K, a gas density of $2\times10^4$ cm$^{-3}$, and a standard cosmic ray ionization rate of $1.3\times10^{-17}$ s$^{-1}$. Initial elemental abundances were taken from \citet{hincelin_oxygen_2011} with the exception of atomic oxygen, where we utilize a slightly carbon rich C/O $\approx1.1$ and $x(O)_{t=0}\approx1.5\times10^{-4}$, as described in \citet{Loomis:2020aa}.

The results of our simulations are shown in Fig. \ref{fig:modeling}. In our network, propargyl cyanide is formed mainly via the reaction

\begin{equation}
    \ce{CN + CH3CCH -> HCCCH2CN + H}.
    \label{r1}
\end{equation}

\noindent
This reaction has, to the best of our knowledge, not been studied in detail. However, based on work by \citet{smith_temperature-dependence_2006}, we assume it occurs barrierlessly since the difference between the ionization energy of propyne, 10.36 eV \citep{llk_ion_2020}, and the electron affinity of the cyano radical, 3.86 eV \citep{bradforth_photoelectron_1993}, is less than $\sim9$ eV. {Based on our assumption that reaction \eqref{r1} is a barrierless process, we have included it in our network with the single-collision rate coefficient of $3\times10^{-10}$ cm$^3$ s$^{-1}$. }

By a similar line of reasoning, we do not include the analogous reactions involving the propargyl radical, \ce{CH2CCH}, namely,

\begin{equation}
    \ce{HCN + CH2CCH -> HCCCH2CN + H}
    \label{r2}
\end{equation}

\begin{equation}
    \ce{HNC + CH2CCH -> HCCCH2CN + H},
    \label{r3}
\end{equation}

\noindent
since the difference between the ionization energies of the closed shell reactants and the propargyl radical is greater than 9 eV in both cases. {Given the likely presence of activation energies for reactions \eqref{r2} and \eqref{r3}, we did not attempt to quantitatively estimate their effect on the abundance propargyl cyanide, though our assumption here is that the rate coefficients will be much smaller than that of reaction \eqref{r1}.}

An additional grain-surface formation route involving the 1-cyano propargyl radical, \ce{HCCCHCN}, i.e.

\begin{equation}
    \ce{H + HCCCHCN -> HCCCH2CN}
    \label{r4}
\end{equation}

\noindent
was also included. Here, the 1-cyano propargyl radical precursor is formed in the gas and on grains via the reaction of carbon atoms with vinyl cyanide \citep{guo_crossed_2006} - {with the barrierless gas-phase formation route also being assumed to occur at the collision rate of $3\times10^{-10}$ cm$^3$ s$^{-1}$ in our network. For reaction \eqref{r4}, and indeed for all such diffusive surface reactions, rate coefficients were calculated using the formula described in \S2.3 of \citet{ruaud_gas_2016}; however, given the low dust temperatures in our simulations and the small fraction of grain-surface \ce{HCCCH2CN} that is non-thermally desorbed, the overall contribution of \eqref{r4} to the gas-phase abundance of propargyl cyanide was negligible compared with \eqref{r1}. Once formed, gas-phase propargyl cyanide is destroyed via reaction with ions, with rate coefficients calculated using the formula given in \citet{woon_quantum_2009}.}

As one can see in Fig. \ref{fig:modeling}, our calculated abundances of propargyl cyanide qualitatively match the observationally derived values to within a factor of a few. In general, we find that the abundances of species such as propargyl cyanide, as well as other unsaturated species such as the cyanopolyynes - with the exception of \ce{HC11N} \citep{Loomis:2020aa} - are much better reproduced than the more complex, aromatic 1-/2-cyanonaphthalene \citep{McGuire:2020aa}, benzonitrile \citep{Burkhardt:2020aa}, and 1-cyano-cyclopentadiene \citep{McCarthy:2020aa}, all of which are severely underproduced. This striking difference in how well our models reproduce the abundances of these species suggests a fundamental shortcoming in how the chemistry of cyclic molecules, generally, is included in our network, and in particular, that there are additional top-down or bottom-up mechanisms that could efficiently form the aromatic precursors.

Furthermore, it is possible that the formation of these aromatic species relies on the cyclization of long carbon chain species.  Detections of both cyclic species and their potential cyclization precursors are therefore important to providing observational constraints on the efficacy of this process. As such, the likely imminent detection of more partially-saturated carbon chains, such as propargyl cyanide, as the GOTHAM survey progresses will provide a fantastic resource of intermediate species for chemical models to compare to when considering a bottom-up (cyclization) formation route for cyclic molecules.

\section{Discussion}
\label{discussion}

The dataset used in these six first-results papers represents only $\sim$30\% of the eventual data that will be collected for the GOTHAM project.  Despite this, and despite covering less total bandwidth, GOTHAM has produced detections of a number of new interstellar molecules that were not seen in the previous work by \citet{Kaifu:2004tk}.  This is true even when applying stacking techniques.  For example, only a small hint of the presence of $c$-\ce{C6H5CN} was seen in stacked data from \citet{Kaifu:2004tk} in the detection presented by \citet{McGuire:2018it}, despite the former work covering more than an order of magnitude more bandwidth.  

The parameters derived for $c$-\ce{C6H5CN} in this analysis are substantially improved over the prior measurements reported in \citet{McGuire:2018it} due to the increased sensitivity and number of observed transitions (See Appendix A).  These new parameters shed some light on why our survey is detecting so many new molecules beyond what is accessible from the \citet{Kaifu:2004tk} survey. We find that the velocity component which accounts for nearly half of the observed column of $c$-\ce{C6H5CN} has a source size of 65$^{+20\prime\prime}_{-13}$.  For comparison, at 25 GHz the GBT HPBW is $\sim$30$^{\prime\prime}$, where as the Nobeyama 45\,m HPBW used for the \citet{Kaifu:2004tk} work is $\sim$67$^{\prime\prime}$.  This would result in a factor of $\sim$2 difference in the line intensity of this component due to beam dilution, and is likely the main reason why $c$-\ce{C6H5CN} was not observed in the \citet{Kaifu:2004tk} survey.

Based on this simple geometric argument, and the range of source sizes we are finding, the observed intensities of rotational lines of other nitrile molecules are expected to be 2-5 times higher with the GBT than if they had been observed with the Nobeyama telescope. Detection of lines from rare isotopic ($^{13}$C and $^{15}$N) species of \ce{HC5N} and \ce{HC7N} in the same observation where $c$-\ce{C6H5CN} was found provides additional evidence for the advantage of a larger telescope \citep{Burkhardt:2018ka}. A further increase in sensitivity was achieved by observing at spectral resolution that is appropriate in this narrow line source. The resolution in the Nobeyama survey of 0.22--1.26\,km s$^{-1}$ was frequently a factor of 2-4 times too low for the very sharp spectral lines in TMC--1 (0.1--0.3\,km~s$^{-1}$ FWHM).  When combined with the sensitivity of the GOTHAM observations, we expect that, in general, our detection limits should be roughly an order of magnitude better than those of \citet{Kaifu:2004tk}. 

Because the observations span a factor of $\sim$3 in frequency (8--29\,GHz), they also span a factor of $\sim$3 in GBT beam size.  As a result, for species with transitions observed across that range of frequencies and beam sizes, such as benzonitrile and the cyanopolyynes discussed above, the effects of beam dilution can be modeled and the effective source sizes used as parameters in the fits.  Still, the source sizes are often the dominant sources of error in our MCMC fits \citep{Loomis:2020aa}.   This is seen quite strongly in the detection of propargyl cyanide presented here.  Because most of the flux from this molecule is concentrated in our K-band observations, the source sizes are relatively unconstrained (Table~\ref{propargylcyanide_results} and Figure~\ref{propargylcyanide_corner}). Observing the source structure directly would substantially improve the certainty in our measurements of column densities and excitation temperatures.

Unfortunately, preliminary efforts to do so have proven difficult.  The emission is too extended to achieve meaningful constraints with GBT maps at frequency $<$50\,GHz.  Early indications are that proof-of-concept observations with the VLA in its most compact D-configuration are proving difficult to calibrate and will lack the surface brightness sensitivity to robustly constrain these source sizes even for bright cyanopolyyne features.  In the short term, the most profitable avenue is likely to be maps of cyanopolyynes in W-band ($>$85\,GHz) with the GBT, providing resolutions of $\sim$10$^{\prime\prime}$.  Longer term, the central core of the Next-Generation VLA (ngVLA) may provide the needed sensitivity and resolution at lower frequencies \citep{Selina:2018uf}.

\section{Conclusions}
\label{conclusions}

Motivated by the detection of $c$-\ce{C6H5CN} in TMC-1, we have begun a large-scale effort to conduct a high-resolution, high-sensitivity spectral line survey of the source using the GBT.  We have presented here an overview of the survey and details of the data reduction procedure.  The dataset shown here represents only $\sim$30\% of the total survey.  The detection of propargyl cyanide (\ce{HCCCH2CN}) for the first time in the ISM is presented here.  Based on our astrochemical modeling, the presence of propargyl cyanide suggests several additional unsaturated --CN containing hydrocarbons may be good targets for interstellar detection.  Using a combination of MCMC fitting techniques and matched filtering algorithms, an additional five new species are also seen from this survey.  We expect to detect several additional new molecules from the completed set of observations.   The dominant source of uncertainty is the underlying source size structure.

\acknowledgments

The authors thank the anonymous referees for their comments, which have improved the quality of this work.  The authors thank D. Frayer, J. Skipper, A. Bonsall, and the staff of the Green Bank Observatory for support with observations, data reduction, and calibration strategies.  A.M.B. acknowledges support from the Smithsonian Institution as a Submillimeter Array (SMA) Fellow. M.C.M. and K.L.K. Lee acknowledge support from NSF grant AST-1615847 and NASA grant 80NSSC18K0396. Support for B.A.M. was provided by NASA through Hubble Fellowship grant \#HST-HF2-51396 awarded by the Space Telescope Science Institute, which is operated by the Association of Universities for Research in Astronomy, Inc., for NASA, under contract NAS5-26555. C.N.S. thanks the Alexander von Humboldt Stiftung/Foundation for their generous support, as well as V. Wakelam for use of the \texttt{NAUTILUS} v1.1 code.  C.X. is a Grote Reber Fellow, and support for this work was provided by the NSF through the Grote Reber Fellowship Program administered by Associated Universities, Inc./National Radio Astronomy Observatory and the Virginia Space Grant Consortium.  E.H. thanks the National Science Foundation for support through grant AST 1906489.  M.C.M and K.L.K.L. acknowledge financial support from NSF grants AST-1908576, AST-1615847, and NASA grant 80NSSC18K0396. S.B.C. and M.A.C. were supported by the NASA Astrobiology Institute through the Goddard Center for Astrobiology.  The National Radio Astronomy Observatory is a facility of the National Science Foundation operated under cooperative agreement by Associated Universities, Inc.  The Green Bank Observatory is a facility of the National Science Foundation operated under cooperative agreement by Associated Universities, Inc.

\appendix

\renewcommand{\thefigure}{A\arabic{figure}}
\renewcommand{\thetable}{A\arabic{table}}
\renewcommand{\theequation}{A\arabic{equation}}
\setcounter{figure}{0}
\setcounter{table}{0}
\setcounter{equation}{0}

\section{Laboratory Measurements}
\label{app:}

\begin{figure*}
\centering
\includegraphics[width=\textwidth]{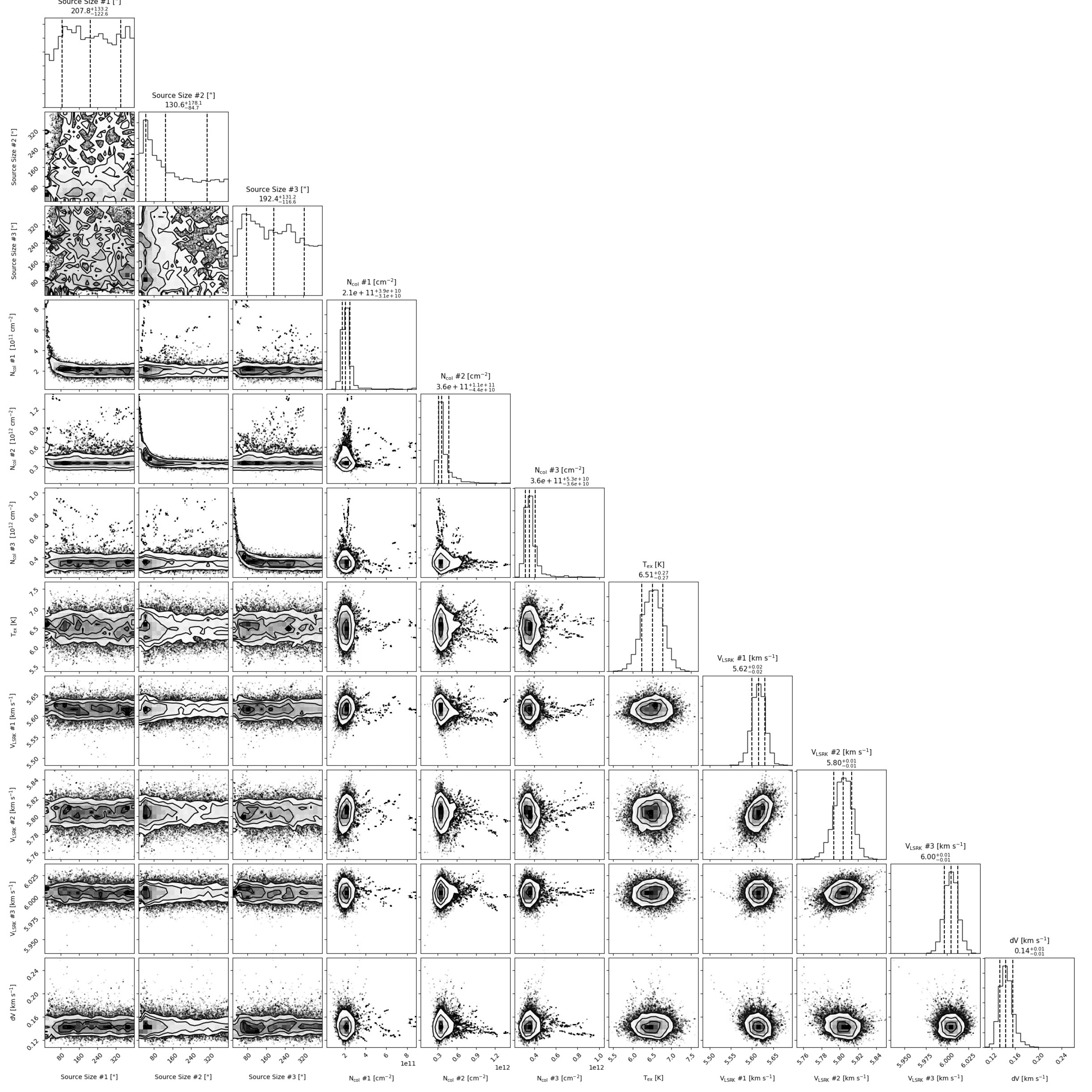}
\caption{Parameter covariances and marginalized posterior distributions for the propargyl cyanide MCMC fit. 16$^{th}$, 50$^{th}$, and 84$^{th}$ confidence intervals (corresponding to $\pm$1 sigma for a Gaussian posterior distribution) are shown as vertical lines. }
\label{propargylcyanide_corner}
\end{figure*}

This appendix includes the newly measured laboratory frequencies of propargyl cyanide (Table~\ref{pcn:lines}), the best-fit spectroscopic constants (Table~\ref{pcn:constants}), the corner plots for propargyl cyanide (Figure~\ref{propargylcyanide_corner}), and the analysis results for benzonitrile (Table~\ref{benzonitrile_results}, Figures~\ref{benzonitrile_spectra},~\ref{benzonitrile_stack}, and~\ref{benzonitrile_corner}).

\begin{table}
\centering
\footnotesize
\caption{Measured Hyperfine-Split  Rotational Transitions of Ground State Propargyl Cyanide}
\label{pcn:lines}
\begin{tabular}{lrrr}
\hline\hline
\multicolumn{2}{c}{Transition}   & Frequency$^1$    & 	Obs.-Calc$^2$  \\
\cline{1-2}
  $J^{\prime}_{K_a,K_c}\rightarrow J^{\prime\prime}_{K_a,K_c}$  &  $F^{\prime} \rightarrow F^{\prime\prime}$  & (MHz)    & 	(MHz)  \\
\hline
$1_{0,1}\rightarrow0_{0,0}$  &  1$\rightarrow$1&  5482.2420(20)&  -0.0014\\
$1_{0,1}\rightarrow0_{0,0}$  &  2$\rightarrow$1&  5482.9248(20)&  0.0004\\
$1_{1,1}\rightarrow2_{0,2}$  &  2$\rightarrow$3&  5949.1197(20)&  -0.0005\\
$2_{1,2}\rightarrow1_{1,1}$  &  2$\rightarrow$1&  10628.9235(20)&  -0.0050\\
$2_{1,2}\rightarrow1_{1,1}$  &  2$\rightarrow$2&  10628.9940(20)&  0.0010\\
$2_{1,2}\rightarrow1_{1,1}$  &  3$\rightarrow$2&  10629.6531(20)&  -0.0003\\
$2_{1,2}\rightarrow1_{1,1}$  &  1$\rightarrow$1&  10629.9537(20)&  -0.0021\\
$2_{1,2}\rightarrow1_{1,1}$  &  1$\rightarrow$0&  10630.1174(20)&  0.0002\\
$2_{0,2}\rightarrow1_{0,1}$  &  2$\rightarrow$2&  10959.9212(20)&  -0.0016\\
$2_{0,2}\rightarrow1_{0,1}$  &  1$\rightarrow$0&  10960.0450(20)&  -0.0002\\
$2_{0,2}\rightarrow1_{0,1}$  &  2$\rightarrow$1&  10960.6057(20)&  0.0020\\
$2_{0,2}\rightarrow1_{0,1}$  &  3$\rightarrow$2&  10960.6587(20)&  0.0006\\
$2_{0,2}\rightarrow1_{0,1}$  &  1$\rightarrow$1&  10961.7465(20)&  -0.0011\\
$2_{1,1}\rightarrow1_{1,0}$  &  2$\rightarrow$1&  11301.6338(20)&  0.0017\\
$2_{1,1}\rightarrow1_{1,0}$  &  1$\rightarrow$1&  11301.7364(20)&  -0.0035\\
$2_{1,1}\rightarrow1_{1,0}$  &  3$\rightarrow$2&  11302.3184(20)&  0.0008\\
$2_{1,1}\rightarrow1_{1,0}$  &  1$\rightarrow$0&  11303.2787(20)&  -0.0020\\
$5_{0,5}\rightarrow4_{1,4}$  &  4$\rightarrow$3&  11854.8251(20)&  0.0026\\
$5_{0,5}\rightarrow4_{1,4}$  &  6$\rightarrow$5&  11854.9046(20)&  0.0000\\
$5_{0,5}\rightarrow4_{1,4}$  &  5$\rightarrow$4&  11855.0503(20)&  -0.0017\\
$3_{1,3}\rightarrow2_{1,2}$  &  3$\rightarrow$2&  15940.9373(20)&  0.0010\\
$3_{1,3}\rightarrow2_{1,2}$  &  2$\rightarrow$1&  15941.0901(20)&  -0.0013\\
$3_{1,3}\rightarrow2_{1,2}$  &  4$\rightarrow$3&  15941.1534(20)&  0.0016\\
$3_{0,3}\rightarrow2_{0,2}$  &  3$\rightarrow$3&  16427.6447(20)&  -0.0013\\
$3_{0,3}\rightarrow2_{0,2}$  &  2$\rightarrow$1&  16428.2776(20)&  -0.0013\\
$3_{0,3}\rightarrow2_{0,2}$  &  3$\rightarrow$2&  16428.3822(20)&  0.0009\\
$3_{0,3}\rightarrow2_{0,2}$  &  4$\rightarrow$3&  16428.4189(20)&  0.0015\\
$3_{0,3}\rightarrow2_{0,2}$  &  2$\rightarrow$2&  16429.4185(20)&  -0.0043\\
$3_{1,2}\rightarrow2_{1,1}$  &  3$\rightarrow$2&  16949.8815(20)&  -0.0016\\
$3_{1,2}\rightarrow2_{1,1}$  &  4$\rightarrow$3&  16950.0758(20)&  -0.0009\\
$3_{1,2}\rightarrow2_{1,1}$  &  2$\rightarrow$1&  16950.1289(20)&  -0.0013\\
$1_{1,0}\rightarrow1_{0,1}$  &  1$\rightarrow$0&  17245.6455(20)&  -0.0001\\
$1_{1,0}\rightarrow1_{0,1}$  &  0$\rightarrow$1&  17245.8069(20)&  -0.0003\\
$1_{1,0}\rightarrow1_{0,1}$  &  2$\rightarrow$2&  17246.0536(20)&  0.0029\\
$1_{1,0}\rightarrow1_{0,1}$  &  1$\rightarrow$1&  17247.3477(20)&  -0.0003\\
$2_{1,1}\rightarrow2_{0,2}$  &  2$\rightarrow$1&  17587.2342(20)&  0.0017\\
$2_{1,1}\rightarrow2_{0,2}$  &  1$\rightarrow$1&  17587.3394(20)&  -0.0009\\
$2_{1,1}\rightarrow2_{0,2}$  &  3$\rightarrow$3&  17587.7128(20)&  0.0025\\
$2_{1,1}\rightarrow2_{0,2}$  &  2$\rightarrow$2&  17588.3772(20)&  0.0008\\
$3_{1,2}\rightarrow3_{0,3}$  &  2$\rightarrow$2&  18109.1948(20)&  0.0032\\
$3_{1,2}\rightarrow3_{0,3}$  &  4$\rightarrow$4&  18109.3717(20)&  0.0021\\
$4_{1,3}\rightarrow4_{0,4}$  &  3$\rightarrow$3&  18821.9361(20)&  0.0008\\
$4_{1,3}\rightarrow4_{0,4}$  &  5$\rightarrow$5&  18822.0522(20)&  0.0007\\
$4_{1,3}\rightarrow4_{0,4}$  &  4$\rightarrow$4&  18822.5011(20)&  -0.0023\\
$4_{1,4}\rightarrow3_{1,3}$  &  4$\rightarrow$3&  21248.8749(20)&  0.0020\\
$4_{1,4}\rightarrow3_{1,3}$  &  3$\rightarrow$2&  21248.9181(20)&  0.0007\\
$4_{1,4}\rightarrow3_{1,3}$  &  5$\rightarrow$4&  21248.9752(20)&  0.0022\\
$4_{0,4}\rightarrow3_{0,3}$  &  3$\rightarrow$2&  21881.1459(20)&  0.0016\\
$4_{0,4}\rightarrow3_{0,3}$  &  4$\rightarrow$3&  21881.1801(20)&  0.0014\\
$4_{0,4}\rightarrow3_{0,3}$  &  5$\rightarrow$4&  21881.2077(20)&  -0.0006\\
$4_{2,3}\rightarrow3_{2,2}$  &  4$\rightarrow$3&  21928.8705(20)&  0.0001\\
$4_{2,3}\rightarrow3_{2,2}$  &  5$\rightarrow$4&  21929.1814(20)&  0.0001\\
$4_{2,3}\rightarrow3_{2,2}$  &  3$\rightarrow$2&  21929.2619(20)&  0.0006\\
$4_{2,2}\rightarrow3_{2,1}$  &  4$\rightarrow$3&  21978.4755(20)&  -0.0022\\
$4_{2,2}\rightarrow3_{2,1}$  &  5$\rightarrow$4&  21978.7757(20)&  -0.0006\\
\hline\hline
\end{tabular}
\end{table}

\setcounter{table}{0}
\begin{table}[t]
\centering
\caption{\textit{Continued} } 
\begin{tabular}{lrrr}
\hline\hline
\multicolumn{2}{c}{Transition}   & Frequency$^1$    & 	Obs.-Calc$^2$  \\
\cline{1-2}
   $J^{\prime}_{K_a,K_c}\rightarrow J^{\prime\prime}_{K_a,K_c}$  &  $F^{\prime} \rightarrow F^{\prime\prime}$  & (MHz)    & 	(MHz)  \\
\hline
$1_{1,1}\rightarrow0_{0,0}$  &  0$\rightarrow$1&  22392.6027(20)&  -0.0031\\
$1_{1,1}\rightarrow0_{0,0}$  &  2$\rightarrow$1&  22392.7013(20)&  -0.0014\\
$1_{1,1}\rightarrow0_{0,0}$  &  1$\rightarrow$1&  22392.7656(20)&  -0.0016\\
$4_{1,3}\rightarrow3_{1,2}$  &  4$\rightarrow$3&  22593.8025(20)&  -0.0015\\
$4_{1,3}\rightarrow3_{1,2}$  &  3$\rightarrow$2&  22593.8892(20)&  0.0012\\
$4_{1,3}\rightarrow3_{1,2}$  &  5$\rightarrow$4&  22593.8892(20)&  -0.0011\\
$7_{0,7}\rightarrow6_{1,6}$  &  6$\rightarrow$5&  24279.3482(20)&  -0.0018\\
$7_{0,7}\rightarrow6_{1,6}$  &  8$\rightarrow$7&  24279.4058(20)&  0.0024\\
$5_{1,5}\rightarrow4_{1,4}$  &  5$\rightarrow$4&  26551.8836(20)&  -0.0021\\
$5_{1,5}\rightarrow4_{1,4}$  &  4$\rightarrow$3&  26551.9079(20)&  0.0047\\
$5_{1,5}\rightarrow4_{1,4}$  &  6$\rightarrow$5&  26551.9460(20)&  0.0024\\
$5_{0,5}\rightarrow4_{0,4}$  &  4$\rightarrow$3&  27314.1358(20)&  -0.0042\\
$5_{0,5}\rightarrow4_{0,4}$  &  5$\rightarrow$4&  27314.1545(20)&  0.0049\\
$5_{0,5}\rightarrow4_{0,4}$  &  6$\rightarrow$5&  27314.1773(20)&  0.0001\\
$5_{2,4}\rightarrow4_{2,3}$  &  6$\rightarrow$5&  27404.8830(20)&  0.0060\\
$5_{2,4}\rightarrow4_{2,3}$  &  4$\rightarrow$3&  27404.8924(20)&  -0.0015\\
$5_{2,3}\rightarrow4_{2,2}$  &  6$\rightarrow$5&  27503.8557(20)&  0.0065\\
$2_{1,2}\rightarrow1_{0,1}$  &  2$\rightarrow$2&  27538.7684(20)&  -0.0029\\
$2_{1,2}\rightarrow1_{0,1}$  &  3$\rightarrow$2&  27539.4316(20)&  -0.0001\\
$2_{1,2}\rightarrow1_{0,1}$  &  2$\rightarrow$1&  27539.4485(20)&  -0.0038\\
$2_{1,2}\rightarrow1_{0,1}$  &  1$\rightarrow$1&  27540.4816(20)&  0.0020\\
$5_{1,4}\rightarrow4_{1,3}$  &  5$\rightarrow$4&  28232.2441(20)&  0.0015\\
$5_{1,4}\rightarrow4_{1,3}$  &  4$\rightarrow$3&  28232.2760(20)&  -0.0064\\
$5_{1,4}\rightarrow4_{1,3}$  &  6$\rightarrow$5&  28232.2937(20)&  0.0025\\
$8_{0,8}\rightarrow7_{1,7}$  &  7$\rightarrow$6&  30590.8950(20)&  -0.0020\\
$8_{0,8}\rightarrow7_{1,7}$  &  9$\rightarrow$8&  30590.9372(20)&  -0.0034\\
$8_{0,8}\rightarrow7_{1,7}$  &  8$\rightarrow$7&  30591.1322(20)&  0.0049\\
$6_{1,6}\rightarrow5_{1,5}$  &  6$\rightarrow$5&  31848.9679(20)&  -0.0024\\
$6_{1,6}\rightarrow5_{1,5}$  &  5$\rightarrow$4&  31848.9850(20)&  0.0057\\
$6_{1,6}\rightarrow5_{1,5}$  &  7$\rightarrow$6&  31849.0099(20)&  0.0011\\
$6_{0,6}\rightarrow5_{0,5}$  &  6$\rightarrow$5&  32722.7003(20)&  0.0004\\
$6_{0,6}\rightarrow5_{0,5}$  &  5$\rightarrow$4&  32722.7003(20)&  -0.0028\\
$6_{0,6}\rightarrow5_{0,5}$  &  7$\rightarrow$6&  32722.7262(20)&  -0.0012\\
$6_{2,5}\rightarrow5_{2,4}$  &  6$\rightarrow$5&  32876.1312(20)&  -0.0030\\
$6_{2,5}\rightarrow5_{2,4}$  &  7$\rightarrow$6&  32876.2340(20)&  0.0008\\
$6_{2,5}\rightarrow5_{2,4}$  &  5$\rightarrow$4&  32876.2340(20)&  0.0006\\
$6_{2,4}\rightarrow5_{2,3}$  &  6$\rightarrow$5&  33048.6853(20)&  -0.0019\\
$6_{2,4}\rightarrow5_{2,3}$  &  5$\rightarrow$4&  33048.7660(20)&  0.0001\\
$6_{2,4}\rightarrow5_{2,3}$  &  7$\rightarrow$6&  33048.7660(20)&  -0.0014\\
$6_{1,5}\rightarrow5_{1,4}$  &  6$\rightarrow$5&  33863.7162(20)&  0.0031\\
$6_{1,5}\rightarrow5_{1,4}$  &  5$\rightarrow$4&  33863.7298(20)&  -0.0069\\
$6_{1,5}\rightarrow5_{1,4}$  &  7$\rightarrow$6&  33863.7493(20)&  0.0042\\
$7_{1,7}\rightarrow6_{1,6}$  &  7$\rightarrow$6&  37139.2049(20)&  0.0027\\
$7_{1,7}\rightarrow6_{1,6}$  &  6$\rightarrow$5&  37139.2049(20)&  -0.0034\\
$7_{1,7}\rightarrow6_{1,6}$  &  8$\rightarrow$7&  37139.2294(20)&  -0.0012\\
$4_{1,4}\rightarrow3_{0,3}$  &  4$\rightarrow$3&  37340.2769(20)&  0.0005\\
$4_{1,4}\rightarrow3_{0,3}$  &  3$\rightarrow$2&  37340.4631(20)&  0.0012\\
\hline\hline
\end{tabular}
\end{table}

\setcounter{table}{0}
\begin{table}[t]
\centering
\caption{\textit{Continued} } 
\begin{tabular}{lrrr}
\hline\hline
\multicolumn{2}{c}{Transition}   & Frequency$^1$    & 	Obs.-Calc$^2$  \\
\cline{1-2}
   $J^{\prime}_{K_a,K_c}\rightarrow J^{\prime\prime}_{K_a,K_c}$  &  $F^{\prime} \rightarrow F^{\prime\prime}$  & (MHz)    & 	(MHz)  \\
\hline
$7_{0,7}\rightarrow6_{0,6}$  &  7$\rightarrow$6&  38102.6913(20)&  -0.0042\\
$7_{0,7}\rightarrow6_{0,6}$  &  8$\rightarrow$7&  38102.7255(20)&  0.0016\\
$7_{2,6}\rightarrow6_{2,5}$  &  7$\rightarrow$6&  38342.3108(20)&  -0.0020\\
$7_{2,6}\rightarrow6_{2,5}$  &  6$\rightarrow$5&  38342.3754(20)&  0.0017\\
$7_{2,6}\rightarrow6_{2,5}$  &  8$\rightarrow$7&  38342.3754(20)&  -0.0030\\
$7_{2,5}\rightarrow6_{2,4}$  &  7$\rightarrow$6&  38616.6908(20)&  -0.0008\\
$7_{2,5}\rightarrow6_{2,4}$  &  6$\rightarrow$5&  38616.7262(20)&  -0.0031\\
$7_{2,5}\rightarrow6_{2,4}$  &  8$\rightarrow$7&  38616.7369(20)&  0.0014\\
$7_{1,6}\rightarrow6_{1,5}$  &  7$\rightarrow$6&  39486.5877(20)&  0.0031\\
$7_{1,6}\rightarrow6_{1,5}$  &  8$\rightarrow$7&  39486.6108(20)&  0.0024\\
\hline\hline
\end{tabular}
\\
$^{1}$ Estimated experimental uncertainties (1$\sigma$) are in units of the last significant digit.\\
$^{2}$ Calculated frequencies derived from the best-fit constants listed in Table~\ref{pcn:constants}.\\
\end{table}

\begin{table}
\centering
\caption{Best-fit Spectroscopic Constants of Ground State Propargyl Cyanide}
\label{pcn:constants}
\begin{tabular}{lrrr}
\hline\hline
Constant        	  &  \citet{demaison_millimeter-wave_1985}$^1$ & This work$^{1,3}$   \\
\hline
$A$ (MHz)            &	 19820.1789(22)	&	19820.080(70) 			  \\
$B$ (MHz)            &	2909.59089(33)	    &	2909.6062(12)  		  \\
$C$ (MHz)            &	2573.22455(35)	    &	2573.2123(12)   		  \\
\\					
$\Delta_J$ (kHz)     &	1.86972(15)		    &	1.9046(14)   			  \\
$\Delta_{JK}$ (kHz)  & -67.6822(17)		    &	-67.911(22)    				  \\
$\Delta_K$ (kHz)     &	832.439(29)		    &	0.722(69)   			  \\
$\delta_J$ (kHz)     &	0.521828(56)		    & 0.52354(82)   		  \\
$\delta_K$ (kHz)     &		$\cdots$		&	6.35(58)   		  \\
\\		
$\chi_{aa}$(N) (MHz) &	-2.2558(22)$^2$         &	-2.2699(14)    				  \\
$\chi_{bb}$(N) (MHz) &	  0.2102(29)        &     0.2154(15)  				  \\
\\
Number of hfs components &      28          &     112                 \\
$\sigma$ (MHz)       &    0.0017            &     0.0025              \\
weighted average     &    0.847             &     1.233               \\
\hline\hline
\end{tabular}
\\
$^1$ Uncertainties (1$\sigma$) are in units of the last significant digit.\\
$^2$ Hyperfine parameters are from \citet{Jager:1990ty}.\\
$^{3}$ Derived from the measurements in Table~\ref{pcn:lines}.\\ 
\end{table}

\clearpage

\subsection{Propargyl Cyanide Results}

\clearpage

\subsection{Benzonitrile Results}

\begin{figure*}[!t]
    \centering
    \includegraphics[width=\textwidth]{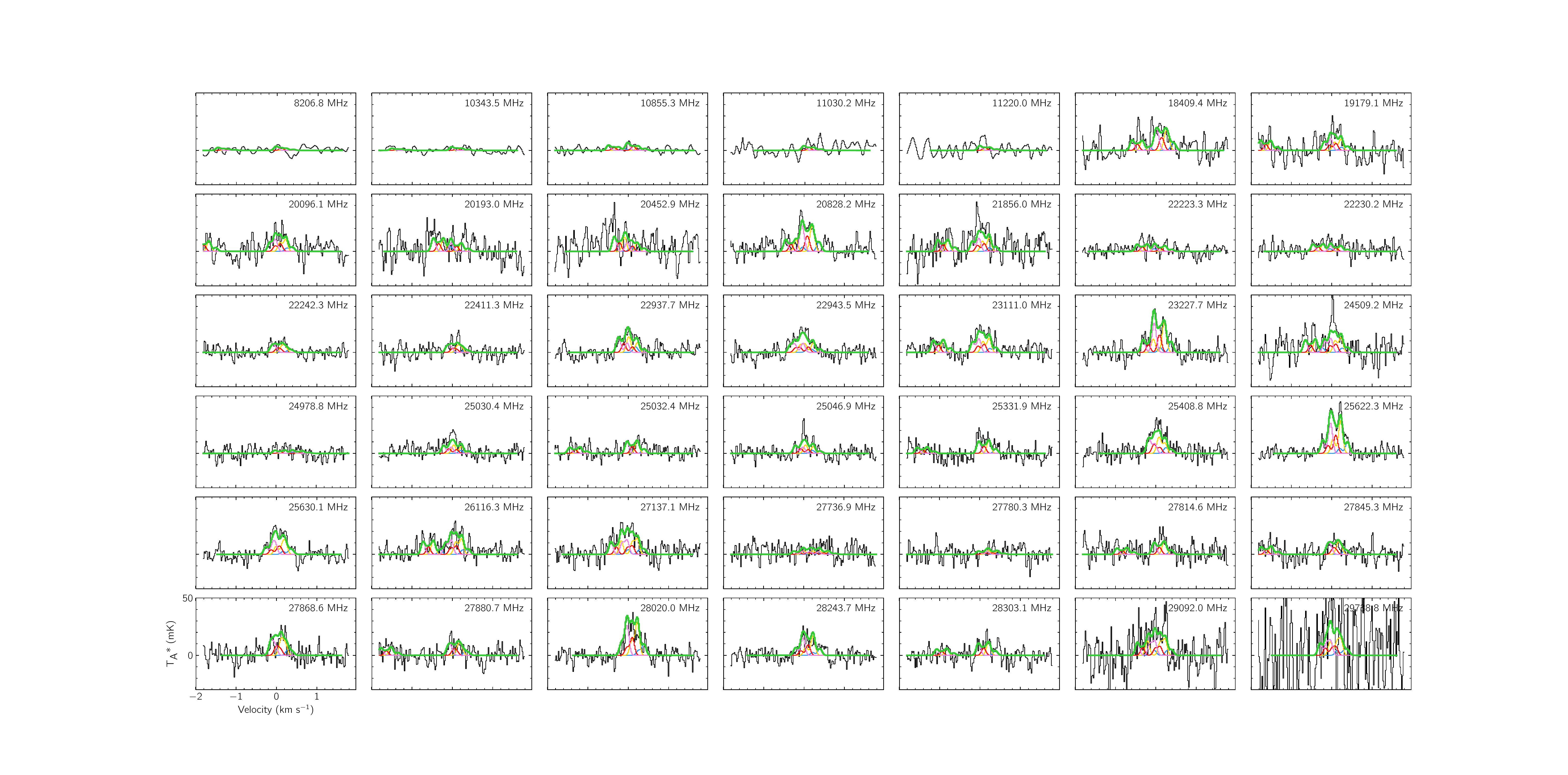}
    \caption{Individual line detections of $c$-\ce{C6H5CN} in the GOTHAM data.  The spectra (black) are displayed in velocity space relative to 5.8\,km\,s$^{-1}$, and using the rest frequency given in the top right of each panel.  The best-fit model to the data, including all velocity components, is overlaid in green.  Simulated spectra of the individual velocity components are shown in: blue (5.595\,km\,s$^{-1}$), gold (5.764\,km\,s$^{-1}$), red (5.886\,km\,s$^{-1}$), and violet (6.017\,km\,s$^{-1}$).  See Table~\ref{benzonitrile_results}.}
    \label{benzonitrile_spectra}
\end{figure*}

\begin{figure*}
    \centering
    \includegraphics[width=0.49\textwidth]{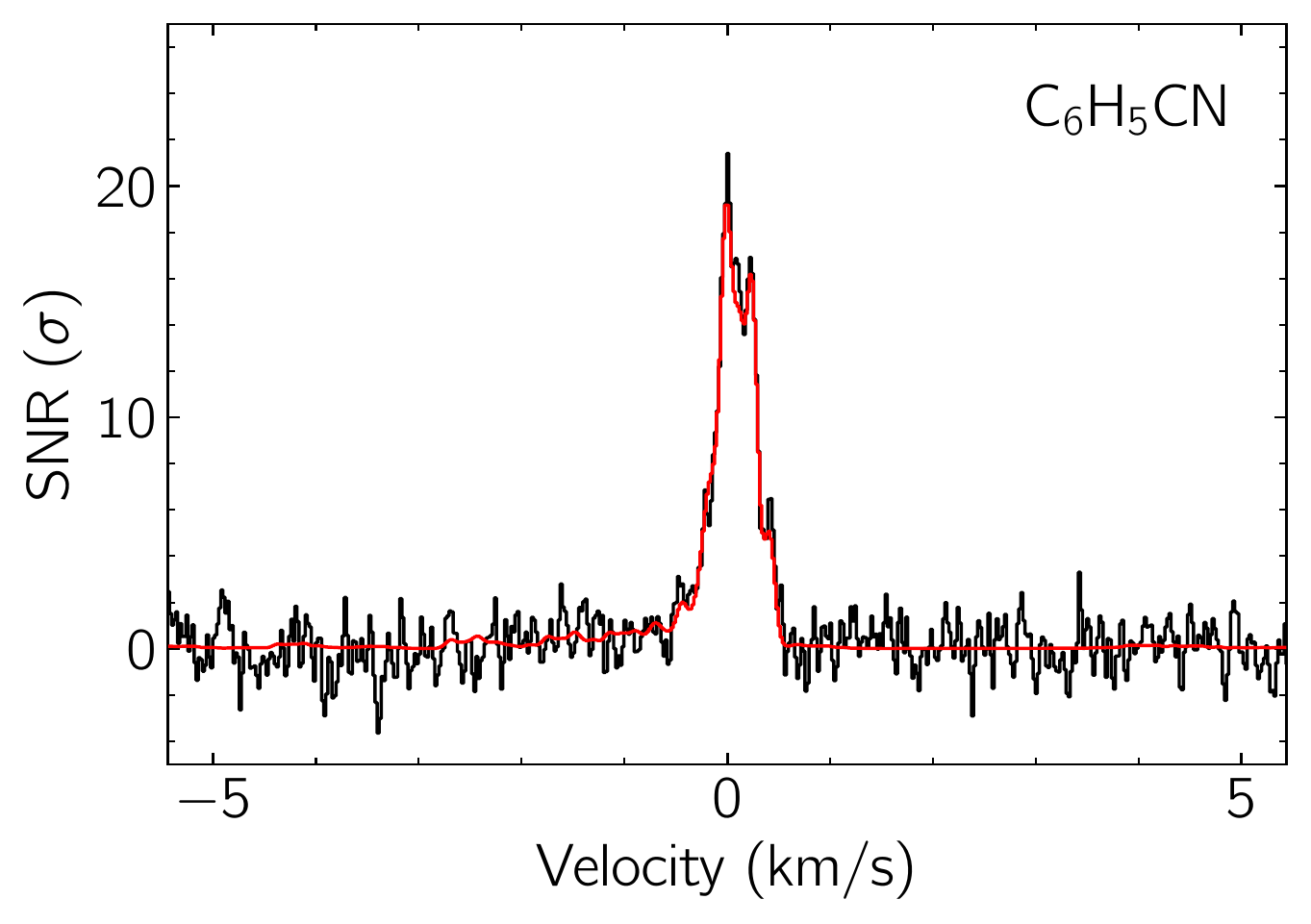}
    \includegraphics[width=0.49\textwidth]{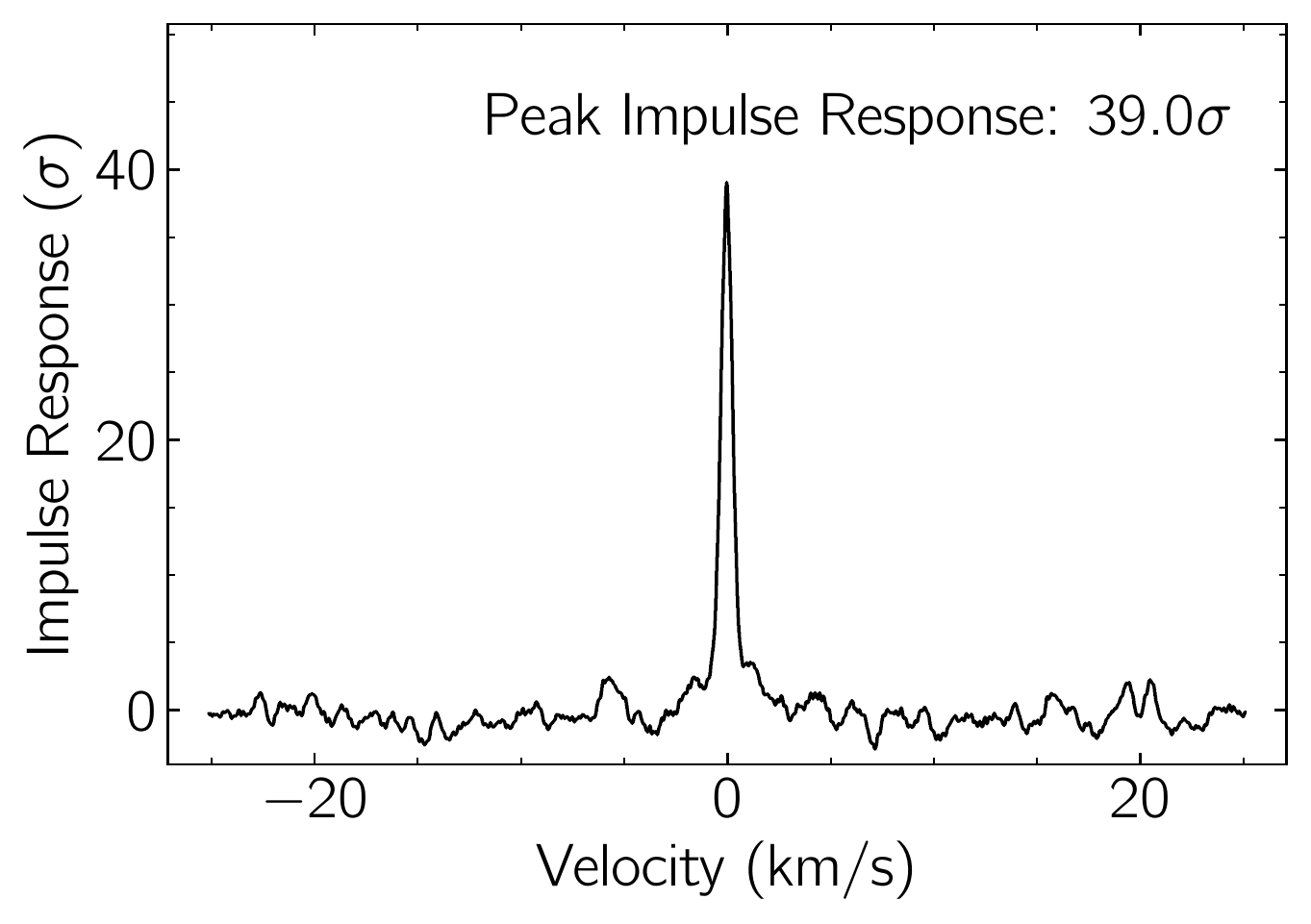}
    \caption{\emph{Left:} Velocity-stacked spectra of $c$-\ce{C6H5CN} in black, with the corresponding stack of the simulation using the best-fit parameters to the individual lines in red.  The data have been uniformly sampled to a resolution of 0.02\,km\,s$^{-1}$.  The intensity scale is the signal-to-noise ratio of the spectrum at any given velocity. \emph{Right:} Impulse response function of the stacked spectrum using the simulated line profile as a matched filter.  The intensity scale is the signal-to-noise ratio of the response function when centered at a given velocity.  The peak of the impulse response function provides a minimum significance for the detection of 39.0$\sigma$.}
    \label{benzonitrile_stack}
\end{figure*}

\begin{figure*}
\centering
\includegraphics[width=\textwidth]{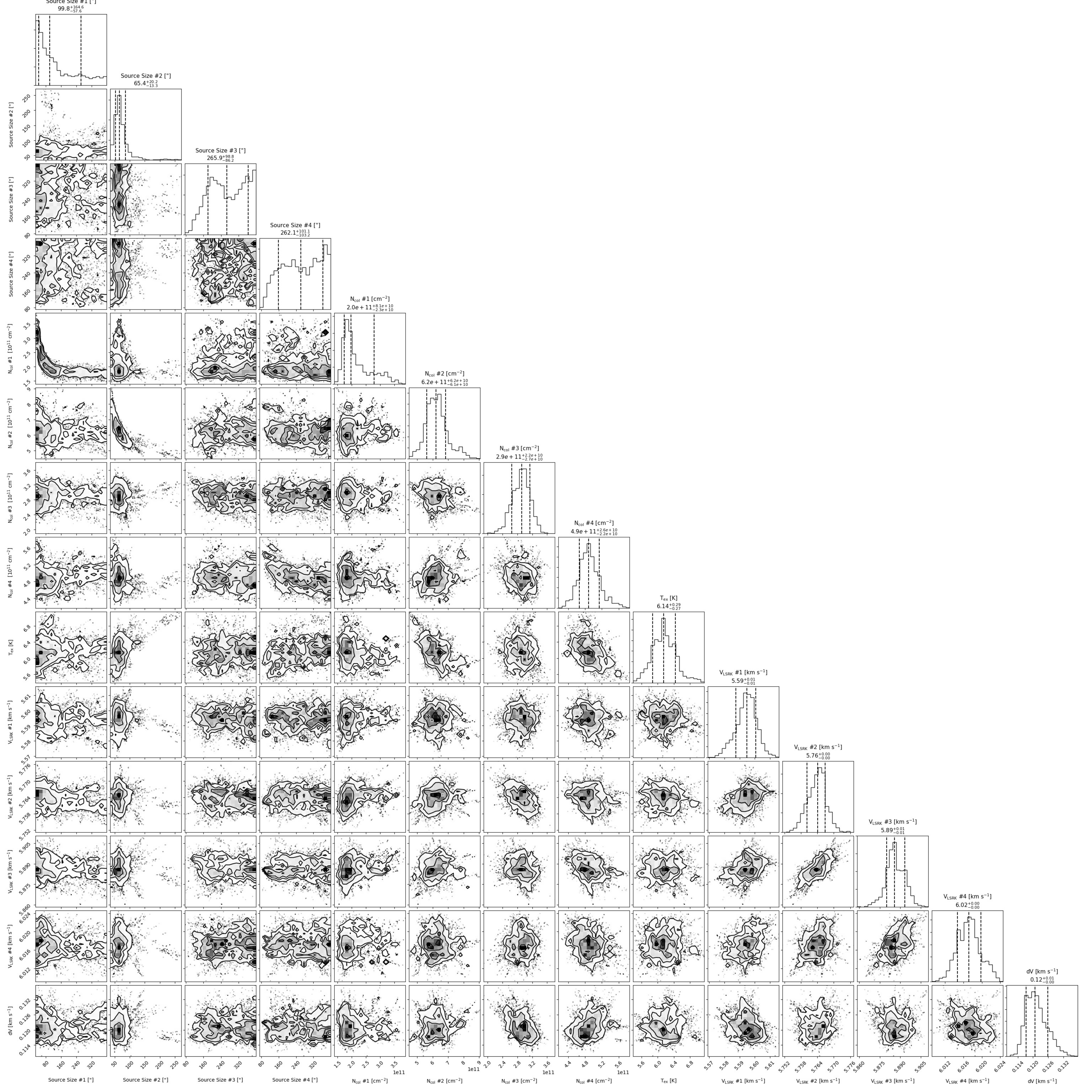}
\caption{Parameter covariances and marginalized posterior distributions for the $c$-\ce{C6H5CN} MCMC fit. 16$^{th}$, 50$^{th}$, and 84$^{th}$ confidence intervals (corresponding to $\pm$1 sigma for a Gaussian posterior distribution) are shown as vertical lines. }
\label{benzonitrile_corner}
\end{figure*}

The results of our MCMC fit to the dataset for $c$-\ce{C6H5CN} are provided below, and are substantially more robust than those achieved in the initial detection from \citet{McGuire:2018it}.  The best-fit parameters are given in Table~\ref{benzonitrile_results} with the associated corner plot shown in Figure~\ref{benzonitrile_corner}.  The individually detected lines are shown in Table~\ref{benzonitrile_spectra}.  The stacked detection and matched-filter response are shown in Figure~\ref{benzonitrile_stack}.

\begin{table*}
\centering
\caption{\ce{Benzonitrile} best-fit parameters from MCMC analysis}
\begin{tabular}{c c c c c c}
\toprule
\multirow{2}{*}{Component}&	$v_{lsr}$					&	Size					&	\multicolumn{1}{c}{$N_T^\dagger$}					&	$T_{ex}$							&	$\Delta V$		\\
			&	(km s$^{-1}$)				&	($^{\prime\prime}$)		&	\multicolumn{1}{c}{(10$^{11}$ cm$^{-2}$)}		&	(K)								&	(km s$^{-1}$)	\\
\midrule
\hspace{0.1em}\vspace{-0.5em}\\
C1	&	$5.595^{+0.006}_{-0.007}$	&	$99^{+164}_{-57}$	&	$1.98^{+0.81}_{-0.23}$	&	\multirow{6}{*}{$6.1^{+0.3}_{-0.3}$}	&	\multirow{6}{*}{$0.121^{+0.005}_{-0.004}$}\\
\hspace{0.1em}\vspace{-0.5em}\\
C2	&	$5.764^{+0.003}_{-0.004}$	&	$65^{+20}_{-13}$	&	$6.22^{+0.62}_{-0.61}$	&		&	\\
\hspace{0.1em}\vspace{-0.5em}\\
C3	&	$5.886^{+0.007}_{-0.006}$	&	$265^{+98}_{-86}$	&	$2.92^{+0.22}_{-0.27}$	&		&	\\
\hspace{0.1em}\vspace{-0.5em}\\
C4	&	$6.017^{+0.003}_{-0.002}$	&	$262^{+101}_{-103}$	&	$4.88^{+0.26}_{-0.22}$	&		&	\\
\hspace{0.1em}\vspace{-0.5em}\\
\midrule
$N_T$ (Total)$^{\dagger\dagger}$	&	 \multicolumn{5}{c}{$1.60^{+0.11}_{-0.07}\times 10^{12}$~cm$^{-2}$}\\
\bottomrule
\end{tabular}

\begin{minipage}{0.75\textwidth}
	\footnotesize
	{Note} -- The quoted uncertainties represent the 16$^{th}$ and 84$^{th}$ percentile ($1\sigma$ for a Gaussian distribution) uncertainties.\\
	$^\dagger$Column density values are highly covariant with the derived source sizes.  The marginalized uncertainties on the column densities are therefore dominated by the largely unconstrained nature of the source sizes, and not by the signal-to-noise of the observations.  See Fig.~\ref{benzonitrile_corner} for a covariance plot, and \citealt{Loomis:2020aa} for a detailed explanation of the methods used to constrain these quantities and derive the uncertainties.\\
	$^{\dagger\dagger}$Uncertainties derived by adding the uncertainties of the individual components in quadrature.
\end{minipage}

\label{benzonitrile_results}

\end{table*}

\section{MCMC Fitting Overview}
\label{app:lines}

{A total of 68 transitions (including hyperfine components) of propargyl cyanide were covered by GOTHAM observations at the time of analysis and were above our predicted flux threshold of 5\%, as discussed in \citet{Loomis:2020aa}. Of these transitions, none were coincident with interfering transitions of other species, and thus a total of 68 transitions were therefore considered. A total of 156 transitions (including hyperfine components) of benzonitrile were covered by GOTHAM observations at the time of analysis and were above our predicted flux threshold of 5\%, as discussed in \citet{Loomis:2020aa}. Of these transitions, none were coincident with interfering transitions of other species, and thus a total of 156 transitions were therefore considered. For both species, observational data windowed around these transitions, spectroscopic properties of each transition, and the partition function used in the MCMC analysis are provided in the Harvard Dataverse repository \citep{GOTHAMDR1}.\\}

\end{document}